\documentclass[letterpaper]{article} 
\usepackage{aaai17}  
\usepackage{times}  
\usepackage{helvet}  
\usepackage{courier}  
\usepackage{url}  
\usepackage{graphicx}  
\graphicspath{{./figures/}}
\begin{document}
\newcommand{\ce}[1]{\texttt{\small{#1}}}

\title{Hows and Whys of Artificial Intelligence for Public Sector Decisions:\\ Explanation and Evaluation}
\author{Alun Preece \\
Crime and Security Research Institute \\
Cardiff University, UK \\
PreeceAD@cardiff.ac.uk
\And 
Rob Ashelford \\
Y Lab $|$ Nesta \\ 
Cardiff University, UK \\
rob.ashelford@nesta.org.uk
\AND 
Harry Armstrong \\
Nesta \\ 
London, UK \\
harry.armstrong@nesta.org.uk
\And
Dave Braines \\
IBM Research \\ 
Hursley, Hampshire, UK \\
dave\_braines@uk.ibm.com}

\newcommand{\aedit}[1]{{\color{black} #1}}

\maketitle

\begin{abstract}
Evaluation has always been a key challenge in the development of artificial intelligence (AI) based software, due to the technical complexity of the software artifact and, often, its embedding in complex sociotechnical processes. Recent advances in machine learning (ML) enabled by deep neural networks has exacerbated the challenge of evaluating such software due to the opaque nature of these ML-based artifacts. A key related issue is the (in)ability of such systems to generate useful explanations of their outputs, and we argue that the explanation and evaluation problems are closely linked. The paper models the elements of a ML-based AI system in the context of public sector decision (PSD) applications involving both artificial and human intelligence, and maps these elements against issues in both evaluation and explanation, showing how the two are related. We consider a number of common PSD application patterns in the light of our model, and identify a set of key issues connected to explanation and evaluation in each case. Finally, we propose multiple strategies to promote wider adoption of AI/ML technologies in PSD, where each is distinguished by a focus on different elements of our model, allowing PSD policy makers to adopt an approach that best fits their context and concerns.
\end{abstract}

\section{Introduction}

Evaluation is, and has always been, a hard problem in the development of artificial intelligence (AI) based software. For any software system, evaluation comprises verification and validation~\cite{IEEE-STD-610:1990}, defined colloquially as follows:
verification focuses on `building the system right' (i.e., assuring its compliance with technical specifications); 
validation focuses `building the right system' (i.e., assuring the system meets stakeholders' requirements). Both are hard for an AI-based system: verification because the technical complexity of the software artifact resists traditional software verification techniques~\cite{O'Keefe:1993}; validation because AI-based systems commonly operate as part of a complex decision-making system involving both software and human elements, making it very difficult to measure and assure intended effects~\cite{Cummings:2014}. 
Recent years have seen rapid advances in AI software capability, mainly due to breakthroughs in machine learning (ML) using multi-layer (so-called `deep') neural networks~\cite{LeCun:2015}. However, this has further increased the technical complexity of the software artifacts and led to even greater challenges in verification and validation~\cite{Goodfellow:2018}. 

It was recognised early in the widespread development of AI-based software --- during the `knowledge engineering' era of the 1980s~\cite{Buchanan:1984} --- that explanation facilities were closely related to verification and validation because these provide necessary scrutability: (i) to developers of the software system to aid debugging, a key element in verification (`building the system right'); (ii) to end-users to promote trust in the system, a key element in validation (`building the right system'). Early work in explanation for AI-based systems identified two corresponding types of explanation request:\\
\textbf{How?} Requesting a `trace' of the system's working (i.e., `How did you come up with $X$?') \\
\textbf{Why?} Requesting a `rationale' of the system's reasoning (i.e., `Why do you believe $X$?')

A considerable body of research and development in AI software explanation, verification and validation derived from this important distinction, after it was understood that different stakeholders require very different kinds of explanation depending on whether their interest is primarily in verification or validation~\cite{Jackson:1999}. Reminders of this point are still being made today, e.g.,~\cite{Kirsch:2017,Tomsett:2018}. Indeed, explanation has once again emerged as a critical issue in the current ML-driven era of AI, though there is a lack of clarity over terminology~\cite{Lipton:2016}.\footnote{Much of the ML community prefers the term \textit{interpretability} to explainability, though a recent UK Government House of Lords review of AI noted, ``The terminology used by our witnesses varied widely. Many used the term transparency, while others used interpretability or `explainability', sometimes interchangeably. For simplicity, we will use `intelligibility' to refer to the broader issue.''}. Generally, there is some consensus that two approaches exist: (1) \textit{transparency} aims to explain system outputs in terms of the internal workings of the system (i.e., transparency in a technical sense, such as visualisations of neural network activations); (2) \textit{post-hoc explanations} aim to justify outputs in terms of rationalisations of the system's workings rather than trying to show the actual workings (e.g., an `explanation by example' in terms of selecting illustratively-similar previously-seen examples).

In this paper, we focus on AI/ML for public sector decision (PSD) applications where evaluation and explanation are of key importance for several reasons. It can be especially challenging to evaluate the impact of an AI application for PSD because the system will commonly be embedded in a sociotechnical process that may make its effects difficult to measure and validate. Moreover, explanation in such applications is closely connected with issues of accountability~\cite{Diakopoulos:2016} that are particularly acute in the case of publicly-funded bodies and decision processes. Finally, in the vast majority of cases, decisions and actions are executed by human actors in PSD applications, setting high standards for evaluation and explanation in the context of human-machine collaboration~\cite{Terveen:1995}.

The paper is organised as follows: to establish a framework for subsequent discussion, we begin by identifying elements of a human$+$machine AI-based decision loop, and consider how evaluation (verification and validation) and explanation address knowns/unknowns in the loop; we then examine common types of AI applications for PSD, and map these to our framework; then we consider the wider context of AI deployment for PSD in terms of data quality, robustness, human-machine collaboration, asset ownership, and `what works', in each case pin-pointing the key foci in terms of our earlier framework; finally, we propose multiple strategies for AI-driven PSD, again drawing on our framework: `mission-oriented', `data-oriented', `work-oriented', and `evidence-oriented'.

\subsection{Elements of a Human$+$AI Decision Loop}

To frame subsequent discussion, Figure~\ref{fig:loop} shows a conceptual model of a decision cycle involving an ML-based AI system working with a human decision maker. This model is adapted from Boyd's OODA (observe, orient, decide, act) loop~\cite{Boyd:1995}. The main elements of the loop are as follows:

\begin{figure}[t]
\centering
\includegraphics[width=0.47\textwidth]{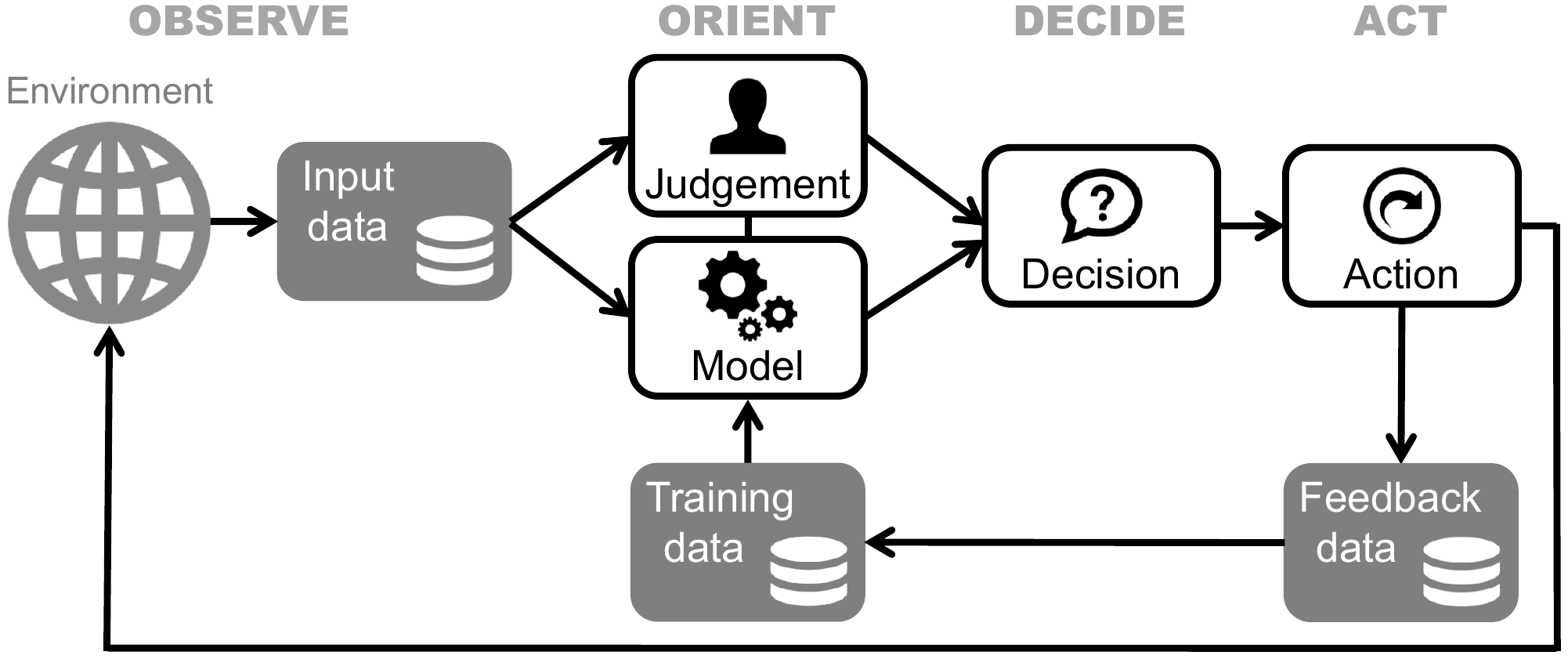}
\caption{Human$+$machine decision loop, based on OODA}
\label{fig:loop}
\end{figure}

\textbf{Input data}: collected from the environment and comprising the observations required as input to the decision-making process. 

\textbf{Model}: the model generated by an ML algorithm from a set of training data; for use in decision-making the model is deployed as part of a software system.

\textbf{Training data}: used to build the model by an ML algorithm, comprising (for supervised learning) a, usually very large, set of input--output pairs describing previous decision cases.

\textbf{Judgement}: for simplicity we refer to the human element of the decision-making process as `judgement', reflecting a combination of knowledge (especially tacit `know how'), experience, and wisdom~\cite{Collins:2010}.

\textbf{Decision}: made in collaboration by human$+$machine at some level of automation from, at one extreme, the machine providing advice to inform a human decision maker to, at the other extreme, the human having some degree of `veto' or supervision over the machine decision maker~\cite{Cummings:2014}.

\textbf{Action}: taken by human(s), machine(s), or a combination of both, that results in feedback and changes the environment.

\textbf{Feedback data}: describing the results of taking the action (may be positive or negative), which may feed back into the set of training data (if the feedback indicates a need for retraining).

The data elements of the model are shown as grey boxes; the operational elements are shown as white boxes.

In terms of OODA, data collection from the environment comprises the \textit{observe} stage and may take the form of physical sensing (e.g., capture of imagery via a camera) or information transfer via documents (e.g., data collected from forms). The ML-based model will operate entirely on this input data as its `view' of the world; the human decision-maker may be able to sense the environment directly as well as accessing the input data (possibly in a form different the that in which the machine accesses it). The multiple models of Boyd's \textit{orient} stage are divided into the ML-based model and the human's judgement. The \textit{decide} and \textit{act} stages are essentially the same here as in OODA, with the difference that Boyd's model allows for continuous feedback from the environment during the decision stage --- which we omit for simplicity --- and our model shows explicit feedback data as an output from the action stage.\footnote{Arguably this data belongs in the `observe' stage since it results from observing the effect of the action, but we feel it useful to place it in the action stage that generates it, allowing it to be fed back into the training data.} The OODA model also shows `guidance and control' between the orient stage and both the observe and act stages, which again we omit for simplicity. 

It has been argued that explanation is a key element in the orient stage of OODA when dealing with goal-directed decision-making~\cite{Aha:2018}. Indeed, we view explanation as an important element in the interaction between human and machine agents in the model (shown by the line connecting those two elements in the figure). In terms of the need for explanations, there are several potential subtleties not explicitly shown in Figure~\ref{fig:loop}, some of which are explored in~\cite{Tomsett:2018}. For example, the human decision maker may not be directly interacting with the AI-based system but instead may be communicating via a operator. In this case, the operator and decision maker may require different forms of explanation. Similarly, humans carrying out the action may require different forms of explanation again. Finally, there may be humans in the environment affected by the action who may have a `right to explanation' as recently enshrined in European law~\cite{Goodman:2016}. We return to this discussion later.

Turning now to evaluation with respect to the elements of Figure~\ref{fig:loop}, we assert that the purpose of evaluation is to consider the space of knowns/unknowns\footnote{As brought to widespread attention in response United States Secretary of Defense Donald Rumsfeld's 2002 quote: https://en.wikipedia.org/wiki/There\_are\_known\_knowns} as follows:

\textbf{Known knowns} are what the AI system creators know the model should know, within the bounds of the training data. These are testable through verification (and retraining where necessary) and are explainable for debugging purposes via transparency type methods.

\textbf{Known unknowns} are queries the AI system creators expect the system to be able to handle, i.e., things that are `predictable' from its training. These are testable through validation and are explainable for user trust purposes via post-hoc explanation methods. The predictions, combined with human judgement, feed forward into decisions and actions, and are ultimately validated via feedback data (possibly requiring retraining).

\textbf{Unknown knowns}, from the perspective of the model, are things outside the scope of the machine, but within the scope of the human decision-maker's knowledge and judgement. Explanations (transparent or post-hoc) may be needed to reveal these machine unknowns. The ability of the human$+$machine ensemble to deal with them must be validated through feedback data, followed potentially by retraining the system in the case of negative feedback.

\textbf{Unknown unknowns} are things outside the scope both of the model and the human's knowledge and judgement. Colloquially, these are ``gotchas''. Robustness of AI-based systems to unknown unknowns has been flagged as a significant current issue, though multiple methods have been identified for mitigating them~\cite{Dietterich:2017}; validation of the human$+$machine ensemble needs to assess the chosen mitigation methods.

\subsection{PSD Problem Types}

Mulgan~\cite{Mulgan:2017} identifies six steps in the PSD process:\\
1) Framing questions for attention;\\
2) Identifying issues that might be amenable to action;\\
3) Generating options to consider;\\
4) Scrutinising/weighing options;\\
5) Deciding (selecting an option);\\
6) Judging whether it worked.

This can be considered both as a decision process and a meta-process for selecting instances of PSD problems. As such, it intersects the decision loop in the preceding section in two ways. Considering it first as a meta-process: Steps~1 to~3 consider which aspects of the environment are `in scope' of the questions, whether appropriate data (input and training) is collectable, whether there are suitable ML algorithms available to generate robust models, and the balance between machine and human intelligence in addressing the questions. Step~4 considers both the pros and cons of ML methods and, importantly, alternative levels of automation between machine and human. Step~5 includes implementation and verification of a chosen approach. Step~6 addresses the broader issues of validation and feedback.

Considering the six steps as a decision process, steps~1 and~2 span observation, steps~2--4 span orientation, step~5 spans decision and action, and step~6 considers feedback. In this view, ML becomes applicable when it can be used to help frame questions, identify issues, and generate options. This is a classic data mining / knowledge discovery type of application and, with this type of application in general, the need to explain the learned model in terms that are meaningful and useful to an end-user (i.e., to transfer the discovered knowledge) is crucial~\cite{Bratko:1997}.

Six `patterns' of PSD data analytics application --- all of which map to the model in the previous section --- have been characterised by the New Orleans Office of Performance and Accountability:\footnote{https://datadriven.nola.gov/nolalytics/}\\
a) `Finding the needle in a haystack':  identify anomalous cases, e.g., by training a predictive model on past anomalous cases;\\
b) `Prioritizing work for impact': classify cases in terms of highest-risk or highest-value; \\
c) `Early warning tools': detect problems at an early stage before escalation, e.g., from a pattern of recurring complaints; \\
d) `Better, quicker decisions': improved decision quality and timeliness driven by maximal use of available data (past cases for training, and input data on the current situation); \\
e) `Optimizing resource allocation': improved organisational efficiency with potential for cost reduction by using means/ends analysis; \\
f) `Experimenting for what works': A/B testing at organisational run-time with dynamic feedback.

For the first three of these, suitable training data is the key element. For the fourth, it is maximising decision quality via human$+$machine collaboration while minimising the `lags' between input data and action, and between action and use of feedback data. The latter two focus on the loop as a whole.

The preceding discussion has implied that the `owner' of the PSD process is a government or public sector body. An alternative view of PSD aims to empower citizens to improve their ability to decide and act in relation to public services and civil society~\cite{Mulgan:2017}.

\subsection{AI for PSD: Issues}

\begin{figure}[t]
\begin{tabular}{p{1.5in}p{1.5in}}
\vspace{0cm}\includegraphics[width=0.2\textwidth]{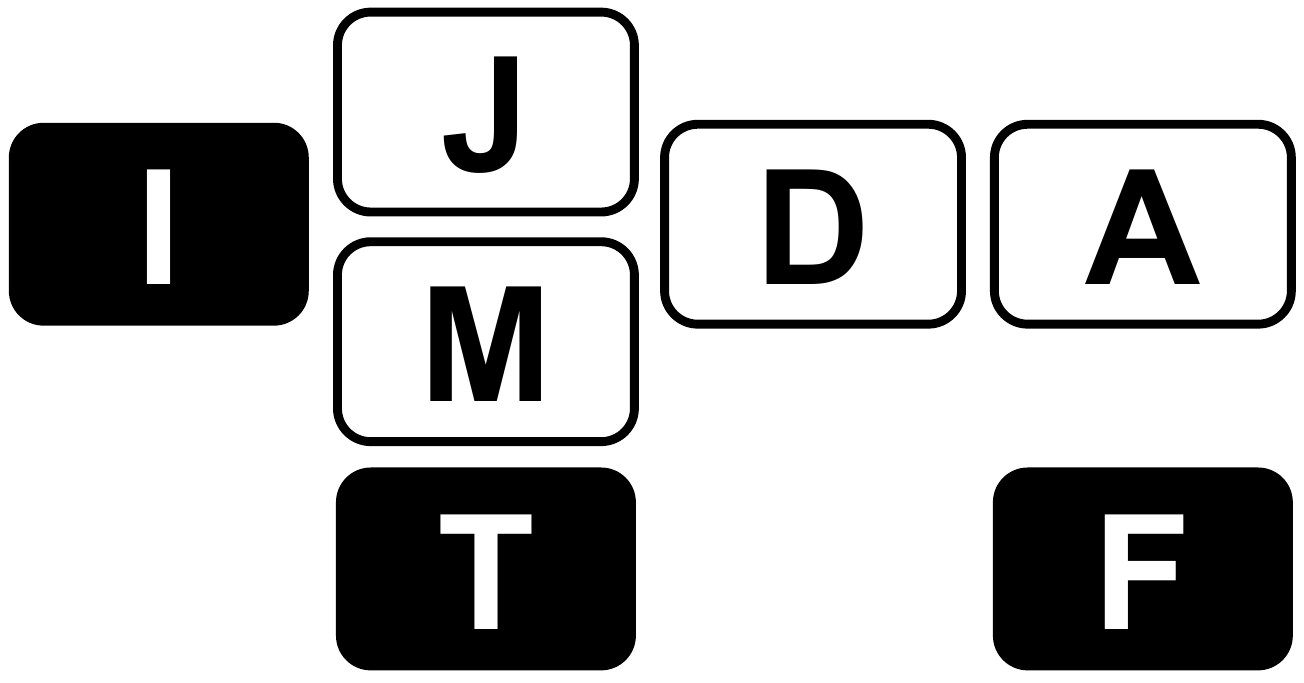} & \vspace{0.2cm}\textbf{Data quality and bias} focuses on the data elements of the model: training, input, feedback.\\ \\
\vspace{0cm}\includegraphics[width=0.2\textwidth]{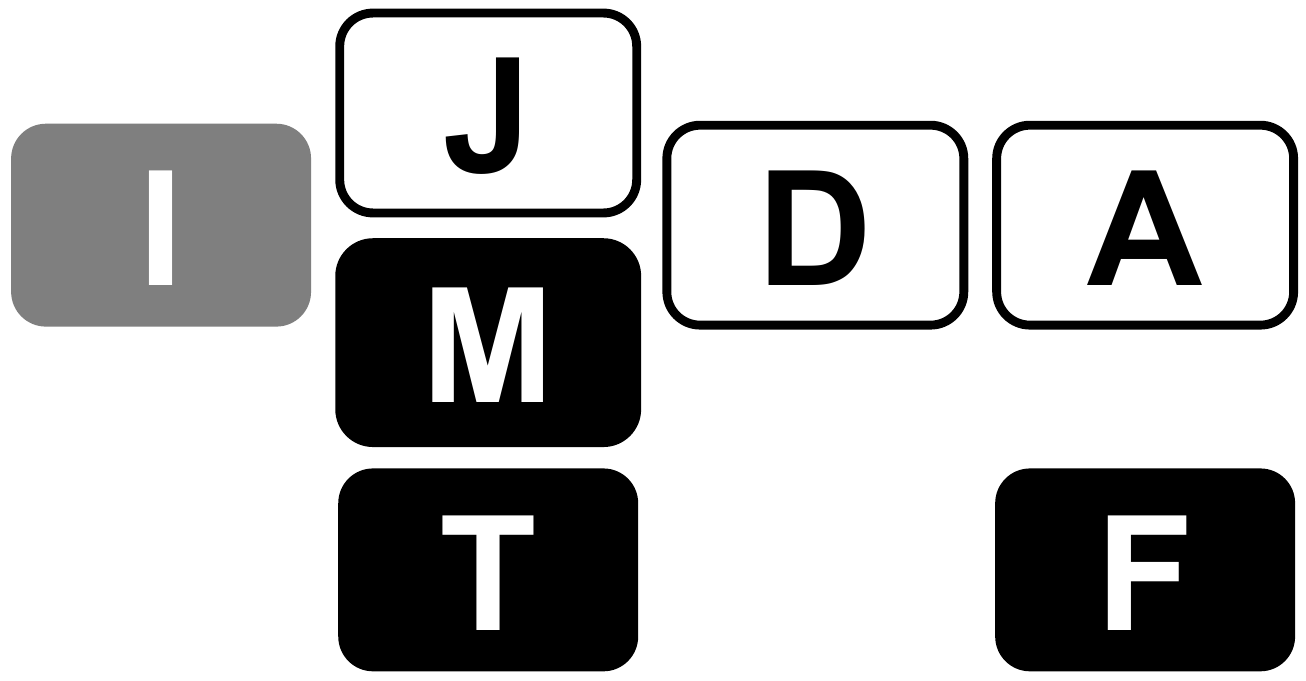} & \vspace{0.2cm}\textbf{Robustness} focuses on the training data, model and feedback data (for retraining).\\ \\
\vspace{0cm}\includegraphics[width=0.2\textwidth]{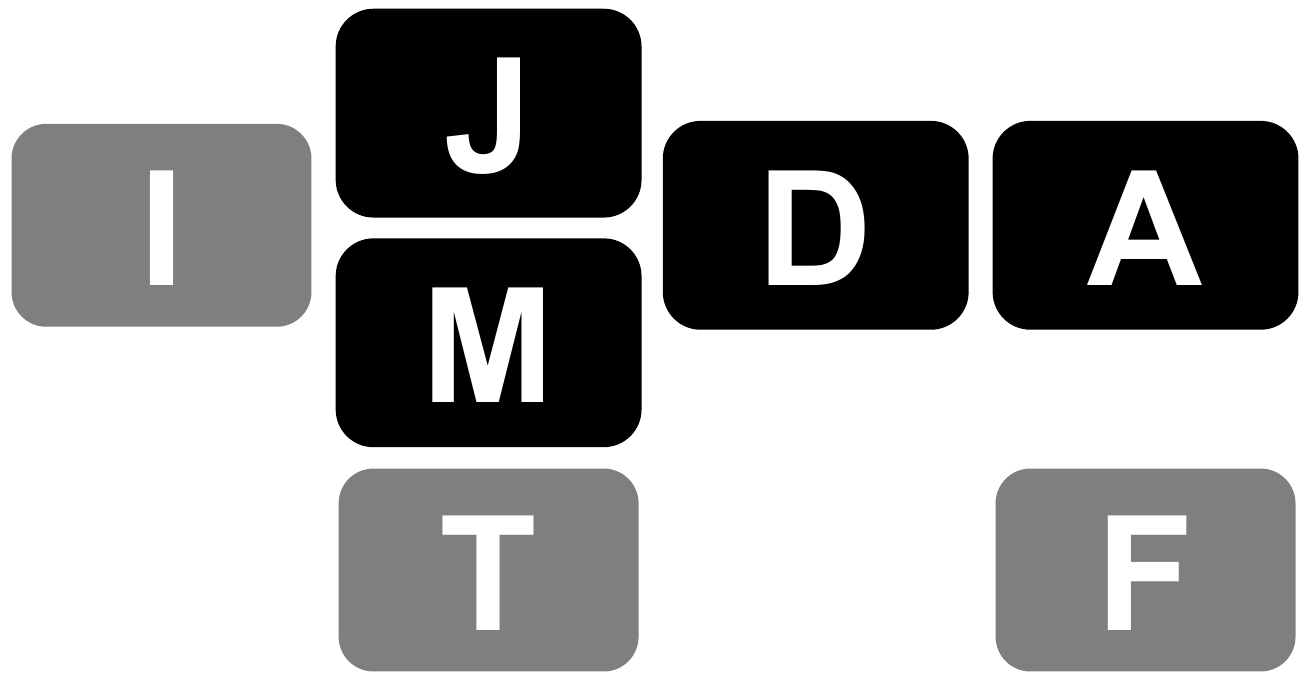} & \vspace{0.2cm}\textbf{Level of Automation} focuses on the relative roles of model and human judgement in decision and action.\\ \\
\vspace{0cm}\includegraphics[width=0.2\textwidth]{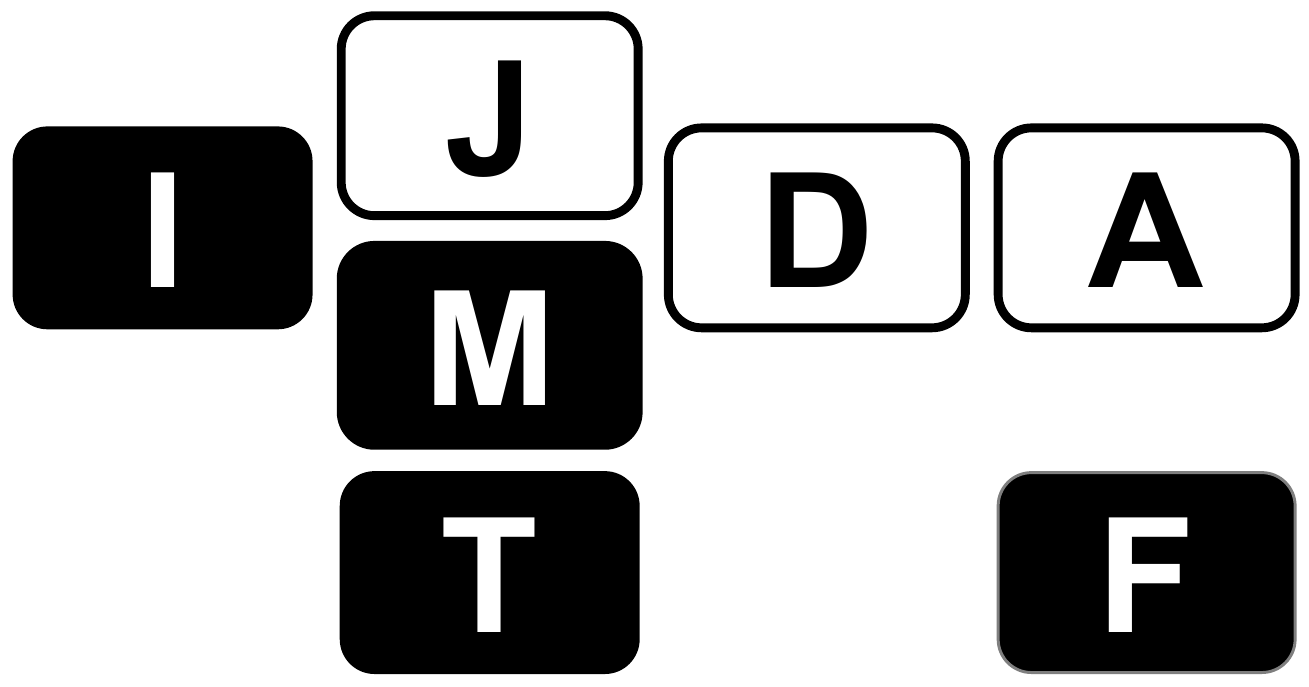} & \vspace{0.2cm}\textbf{Ownership} focuses on the main elements that may be out-sourced: data (training, input, feedback) and model.\\ \\
\vspace{0cm}\includegraphics[width=0.2\textwidth]{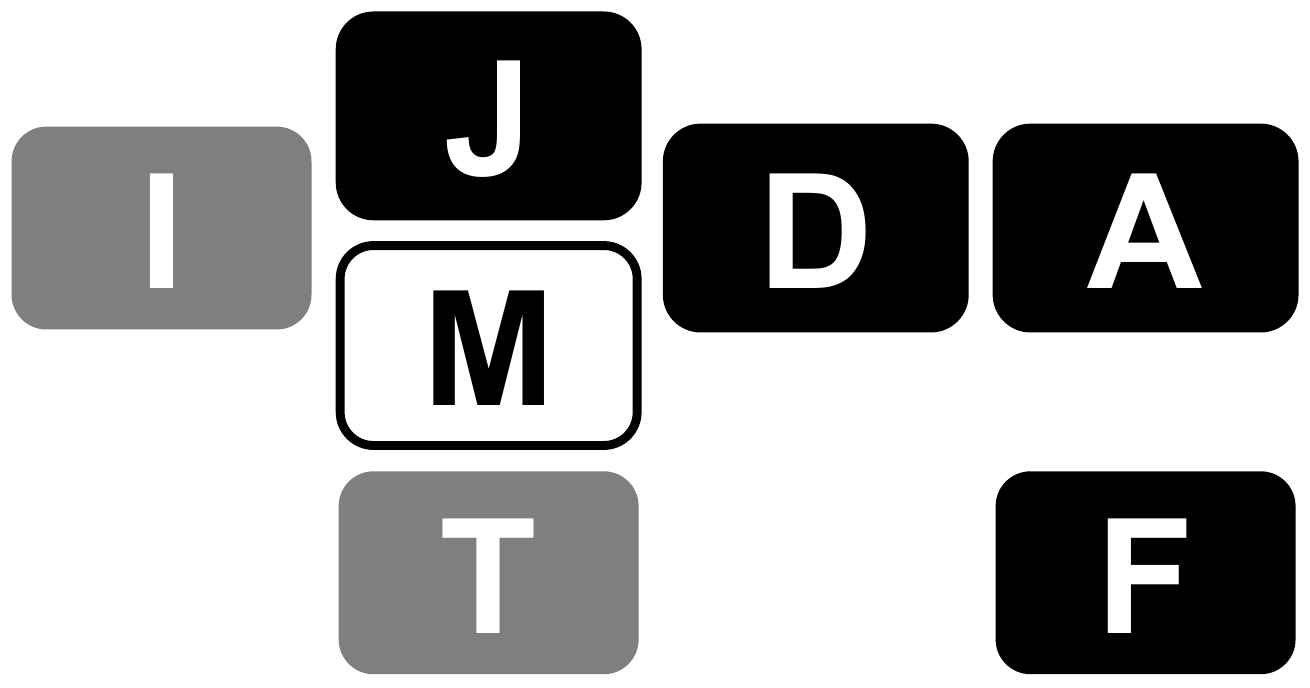} & \textbf{Evaluation and `What Works'} focuses on outcomes: decision, action, feedback data, including the role of human judgement in these.
\end{tabular}
\caption{Main elements of the model (shown in black) in focus for each PSD issue}
\label{fig:issues}
\end{figure}

This section discusses several significant issues requiring attention in the deployment of AI/ML approaches to PSD. 
Figure~\ref{fig:issues} highlights the elements of the loop that are the main focus for each of these issues. In each case, the model is shown as an abstract `periodic table' version of Figure~\ref{fig:loop} with data elements in grey, operational elements in white, and the focus elements (data or operational) in each case shown in black.

\subsubsection{Data Quality and Bias}

Here, the focus is on the data elements of the loop: input, training, feedback. Key considerations include (i) the quality of the training set with respect to scoping known knowns (verification) and known unknowns (validation); (ii) quality of collectable input data (volume, velocity, variety, veracity); (iii) the scope of feedback data including impacts, not just outputs.

\subsubsection{Robustness}

Here, the focus is on model performance --- a function of training data (verification) and feedback (validation) --- and mitigation strategies especially against unknown unknowns~\cite{Dietterich:2017}.

\subsubsection{Level of Automation}

The focus here is on the relative roles of machine vs human in the decision and action. We highlight the importance of explanations (transparent and post-hoc) in run-time validation, e.g., when may the human need to override the machine due to an unknown known. It is also necessary to consider the danger of human overreliance on the machine, making it particularly important that the machine should clearly communicate its own uncertainty.

\subsubsection{Ownership}

In this case, the focus is on potential for outsourcing of key elements of the PSD system, especially  data (input, training, feedback) and model, e.g., contracting an AI vendor to build and manage the system, leaving decision and action in-house. The pros and cons of this must be weighed-up; factors include a lack of AI/ML expertise to do it all in-house, and convenient data management by outsourcing vs retaining value in the data.

\subsubsection{Evaluation and `What Works'}

Here, the focus is on judgement, decisions, actions, and feedback (including impacts): capturing best practices and `what works'. This provides input to using using the model at the start of the previous section as a decision process rather than a meta-process, i.e., to use AI/ML to select promising applications of AI/ML.

\subsection{AI for PSD: Strategies}

\begin{figure}[t]
\begin{tabular}{p{1.5in}p{1.5in}}
\vspace{0cm}\includegraphics[width=0.2\textwidth]{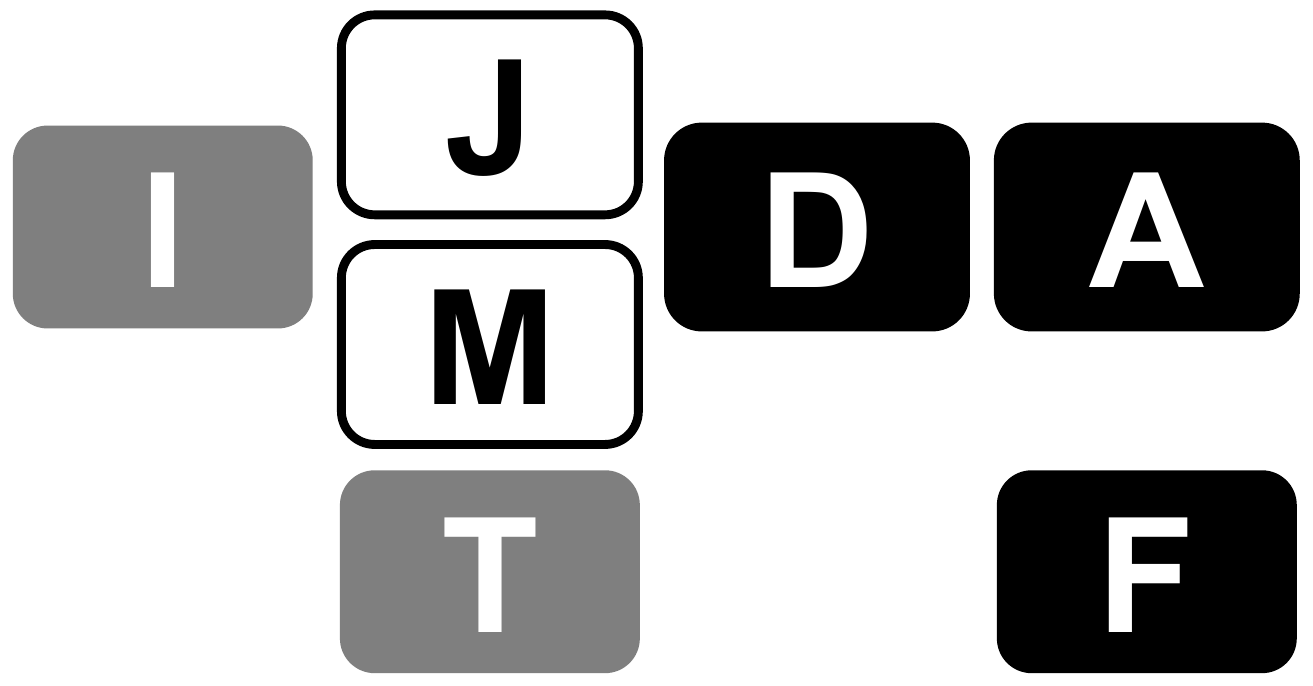} & \vspace{0.2cm}\textbf{Mission-oriented} focuses on outcomes: descision, action and feedback data.\\ \\
\vspace{0cm}\includegraphics[width=0.2\textwidth]{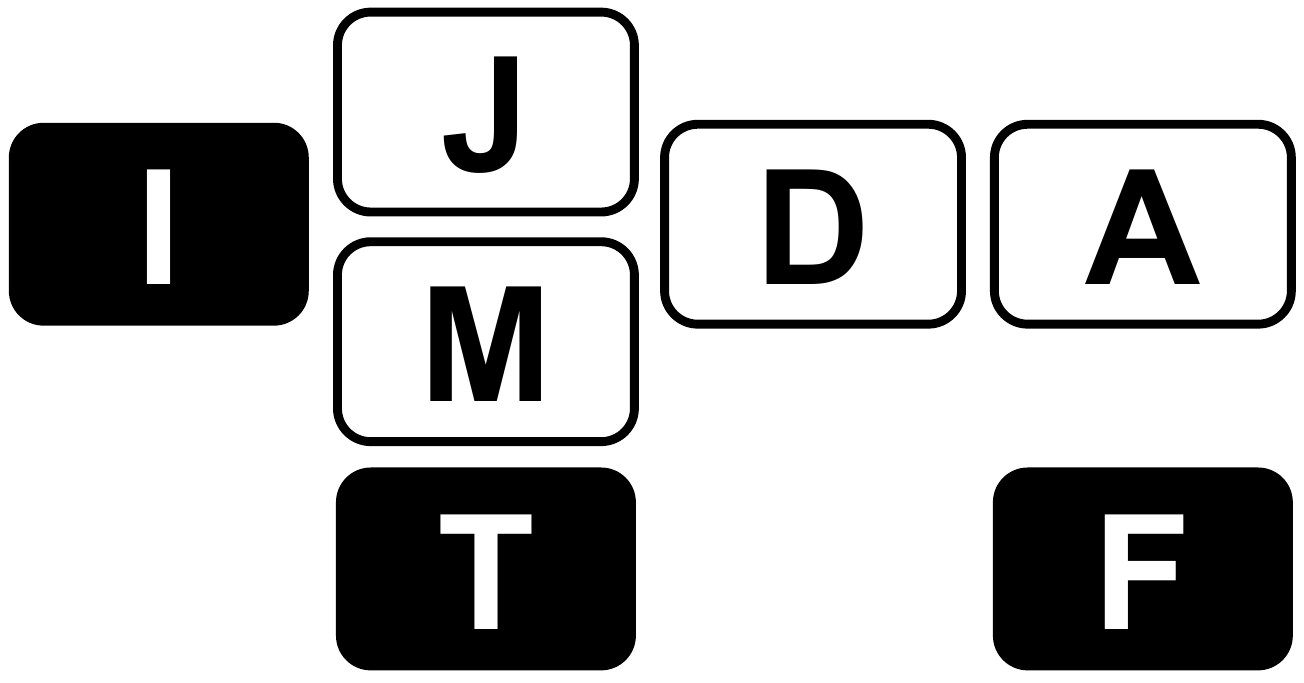} & \vspace{0.2cm}\textbf{Data-oriented} focuses on the data elements: input, training, and feedback.\\ \\
\vspace{0cm}\includegraphics[width=0.2\textwidth]{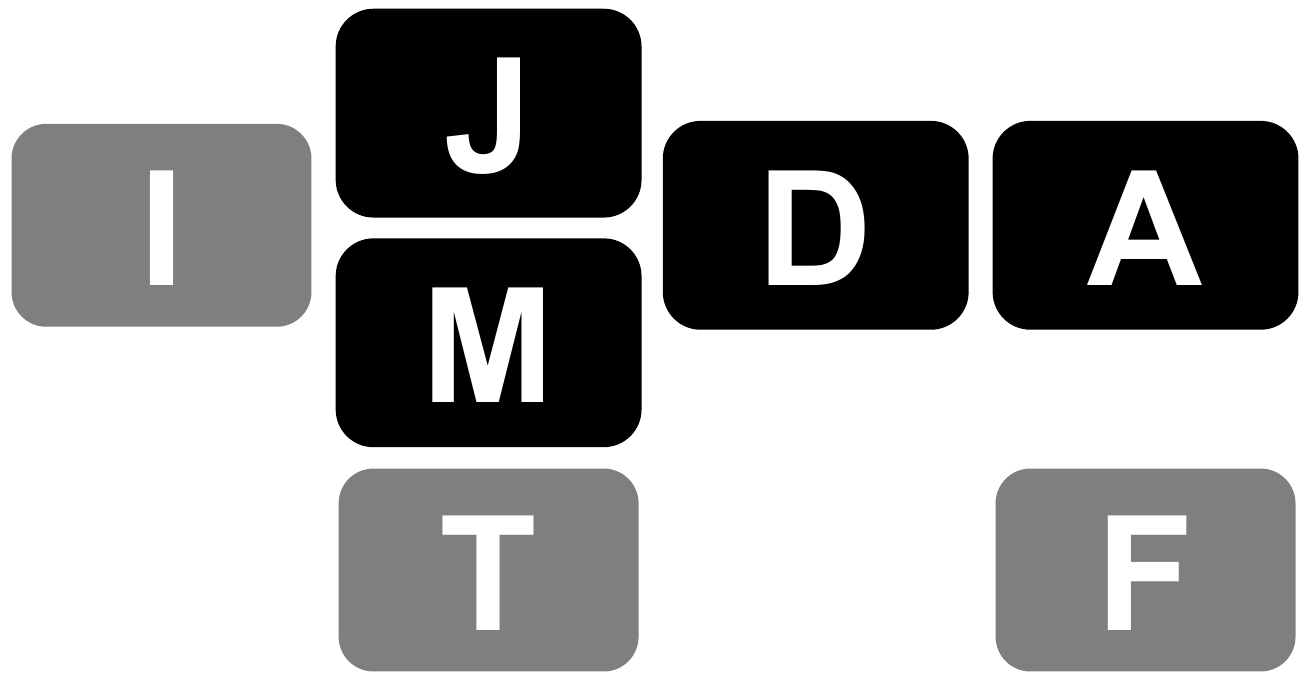} &  \vspace{0.2cm}\textbf{Work-oriented} focuses on the combination of human judgement and ML model in relation to decision and action elements.\\ \\
\vspace{0cm}\includegraphics[width=0.2\textwidth]{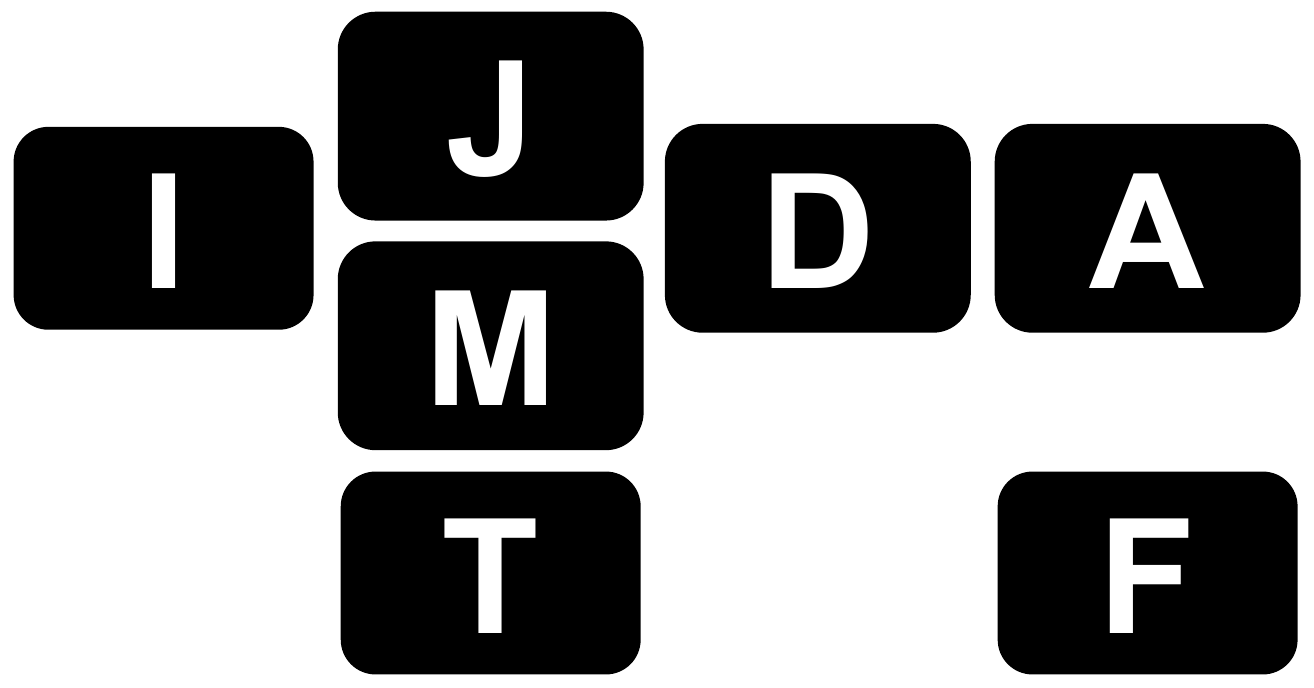} & \vspace{0.2cm}\textbf{Evaluation-oriented} focuses holistically on the entire loop, aiming to capture best practice and `what works'.\\
\end{tabular}
\caption{Main elements of the model (shown in black) in focus for each PSD strategy}
\label{fig:strategies}
\end{figure}

We now turn to examining several strategies to addressing PSD via AI/ML, summarised in Figure~\ref{fig:strategies}.
As before, in each case the model is shown as an abstract version of Figure~\ref{fig:loop} with data elements in grey, operational elements in white, and the focus elements (data or operational) in each case shown in black.

\subsubsection{Mission-oriented}

This strategy aims to maximise the quality of decision-making, and maximise the value from AI/ML to the mission of the public sector organisation. Hence the focus is on outcomes: decision, action and feedback. Explanation probes why what worked worked, and why what didn't work didn't.

\subsubsection{Data-oriented}

Here, the aim is to maximise value from data assets and therefore focuses on the data elements of the model: training, input and feedback. This strategy emphasises effective data management and deployment, either in-house, collectively sector-wide or as public-private partnership. The goal is to increase space of known knowns to greatest extent possible and shift more evaluation burden to verification.

\subsubsection{Work-oriented}

This strategy focuses on maximising value from human and machine collaboration, i.e., the interaction between ML model and human judgement in decision-making and subsequent human/machine roles in action. Thus, the strategy considers carefully the nature of work done by humans and machines in the organisation, with a view to deploying each asset most effectively. Explanation for validation, transparency and trust is key. The strategy yields opportunities for improved systemic robustness (e.g., against unknown knowns and unknown unknowns) and better job satisfaction if done well (human assets are better able to cope with service demands).

\subsubsection{Evaluation-oriented}

This is a holistic strategy: it considers the whole loop with an intent to capture best practice: `what works', including best practice in evaluation and explanation. The aim is to create an evidence base for AI interventions.

\subsection{Discussion and Conclusion}

It is important in system evaluation to distinguish between system outputs and impacts. While traditional machine learning measures such as accuracy and confusion matrices (true/false positives/negatives) are informative of the performance of an ML model, the question of whether the system is having the `right' impact on the organisation is a separate question. The whole loop needs to be considered from a perspective organisational integration and process design. It is rare that an AI system will simply be `dropped into' existing processes/organisations/systems and expected to have the desired impact. Moreover, the above model of human$+$machine decision-making assumes that the human element of decision making in the target organisation is well delineated --- often, in large organisations, decision making is distributed, so understanding the impact of introducing one or more AI elements into such a collective system becomes extremely challenging.

The need for explanation and evaluation must be appropriate for measuring impact. Transparency is not always a significant issue. For example, in many medical interventions, it is unclear how the dynamics work at molecular or systems level (particularly for certain drugs), but we nevertheless have solid evidence to show they work. Legitimate post-hoc explanations may draw on this evidence base. Building the kind of evidence base discussed in the previous section could provide a similar basis for PSD using AI: practitioners could work with trusted bodies that would use the evidence base to recommend AI systems, tools, or approaches. Nevertheless, the feedback element of our model is still important: a `proven' intervention may fail in a novel context due to unknown factors (particularly unknown knowns and unknown unknowns). The feedback improves the AI model and the evidence base. Single, double and triple loop learning is key here~\cite{Mulgan:2017}.

Again, this discussion gets to the key distinction between `building the system right' (high accuracy etc) and `building the right system' (high impact). The `AI Winter' that began in the late 1980s was due to a failure of the technology to meet user expectations despite often high technical performance. The space of viable, impactful applications was much smaller than developers and investors hoped. A solid evidence base for AI impact is needed to challenge the hype that is once again surrounding technological advances in the field.

\end{document}